\documentstyle[12pt,a4,epsf]{article}

\def\preprintno{IKDA 93/28 \\ SI 93-4 \\ hep-ph/9308237}
\makeatletter
\def\@maketitle{\newpage \null
  \@ifundefined{preprintno}
  {\vskip 2em}              
  {                         
    \vspace*{-1\headsep}    
    \vspace*{-1\headheight}
    \begin{flushright}      
      \large \preprintno
    \end{flushright}
    \vspace*{-3em}
    \vskip \headsep	    
    \vskip \headheight
    }                       
  \begin{center}            
    {\LARGE \@title \par} \vskip 1.5em {\large \lineskip .5em
      \begin{tabular}[t]{c}\@author
      \end{tabular}\par}
    \vskip 1em {\large \@date} \end{center}
    \par
    \vskip 1.5em}
\makeatother

\begin{document}

\title{WOPPER -- \\
  A Monte Carlo Event Generator for \\
  {\bf W} {\bf O}ff-shell {\bf P}air {\bf P}roduction \\
  including Higher Order \\
  {\bf E}lectromagnetic {\bf R}adiative Corrections%
  \thanks{Contribution to the Second Workshop -- Munich, Annecy, Hamburg:
    $e^+e^-$ Collisions at 500~GeV: The Physics Potential,
    November 20, 1992, to April 3, 1993}
}
\author{H. Anlauf%
  \thanks{Email: anlauf@crunch.ikp.physik.th-darmstadt.de},
  P. Manakos, T. Mannel%
  \thanks{Email: mannel@crunch.ikp.physik.th-darmstadt.de} \\
  {\it Institut f\"ur Kernphysik, D--64289 Darmstadt, Germany,}
  \vspace*{1mm} \\
  H. D. Dahmen \\
  {\it Universit\"at Siegen, D--57076 Siegen, Germany}
}
\def\today{\relax}

\maketitle

\pagestyle{empty}
\thispagestyle{empty}

\begin{abstract}
  \noindent
  We present the Monte Carlo event generator {\tt WOPPER} for pair
  production of $W$'s and their decays at high energy $e^+e^-$
  colliders.  {\tt WOPPER} includes the effects from finite $W$ width
  and focusses on the calculation of higher order electromagnetic
  corrections in the leading log approximation including soft photon
  exponentiation and explicit generation of exclusive hard photons.
\end{abstract}


\section{Introduction}

The precision experiments being performed at the $e^+e^-$ colliders LEP
at CERN and SLC at SLAC have confirmed the predictions of the Standard
Model (SM) for the interactions between the gauge bosons and the
fermions even at the level of electroweak radiative corrections.
However, the non-Abelian structure of the gauge sector of the SM with
its couplings between the electroweak gauge bosons has not been tested
directly.  In addition, the origin of electroweak symmetry breaking
giving longitudinal components to the electroweak gauge bosons, is still
obscure.

Anomalous couplings will disturb the extensive gauge cancellations
taking place in the SM, and possible new physics will show up in
particular in the cross section for the production of longitudinally
polarized $W$ bosons.  On the other hand, possible new physics is
already severely constrained by LEP100 data, and the effects to be
expected at LEP200 or at a 500~GeV linear collider are small.

In order to extract these small effects, one has to have a precise
knowledge of the radiative corrections within the SM.  The one-loop
electroweak radiative corrections to the production of on-shell $W$'s
are by now well established \cite{RC}.  The influence of the finite $W$
width has been investigated in~\cite{Finite-Width}.  Also, the higher
order QED corrections have been calculated in the leading log
approximation (LLA)~\cite{CDMN91}.

However, the experimental reconstruction of the $W$'s is complicated
by the fact that they may decay either into leptons with an escaping
neutrino, or into hadrons, where the jet energies may be poorly known
due to undetected particles.  In addition, the radiative corrections
due to emission of photons produce a systematic shift of the effective
center of mass energy towards smaller values.  Such effects may best
be studied with the help of a Monte Carlo event generator.

In this note, we report on the status of the new Monte Carlo event
generator {\tt WOPPER} for the process $e^+e^- \to (W^+W^-)^* \to$
4~fermions.  {\tt WOPPER} includes higher order electromagnetic
radiative corrections in the LLA, with explicit generation of exclusive
hard photons and the effects from finite width of the $W$'s.  We present
results from first simulations obtained with this generator, which will
eventually become a full four fermion generator for high energy $e^+e^-$
colliders.


\section{The Monte Carlo generator {\tt WOPPER} }

The Monte Carlo event generator {\tt WOPPER} is capable of a full
simulation of the cross section for $e^+ e^- \to 4$ fermions + $n\gamma$
via the resonant channel containing two $W$ bosons.  The finite width of
the $W$ bosons is included as well as QED radiative corrections in all
orders of the leading logarithmic approximation.  At very high energies
these corrections are indeed the ones which are numerically most
important, since
\begin{equation}
  \frac{\alpha}{\pi} \log \left(\frac{s}{m_e^2}\right)  \approx 6\%
  \qquad  \mbox{(at LEP200 and EE500 energies)}
\end{equation}

The LLA is conveniently incorporated using the so-called structure
function formalism \cite{Structure}.  In this formalism, the expression
for the radiatively corrected cross section reads
\begin{equation}
  \sigma(s) = \int\limits_0^1 dx_+ dx_- \;
  D(x_+,Q^2) D(x_-,Q^2) \; \hat\sigma(x_+ x_- s) \; ,
  \label{eq:factorization}
\end{equation}
where $\hat\sigma$ is the Born level cross section of the hard process,
$D(x,Q^2)$ are the structure functions for initial state radiation, and
$Q^2 \sim s$ is the factorization scale.  The structure function
satisfies the evolution equation
\begin{equation}
\label{eq:DGLAP}
   Q^2 \frac{\partial}{\partial Q^2} D(x,Q^2)
      =  \frac{\alpha}{2\pi}
             \int\limits_x^1 \frac{dz}{z} \left[P_{ee}(z)\right]_+
                 D\left(\frac{x}{z},Q^2\right)
  \quad \mbox{ with } \quad
  P_{ee}(z)  =  \frac{1+z^2}{1-z}
\end{equation}
with initial condition $ D(x,m_e^2) = \delta(1-x) $.  The solution to
eq.~(\ref{eq:DGLAP}) automatically includes the exponentiation of the
soft photon contributions as well as a resummation of the large
logarithms of the form $\log (s/m_e^2)$ from multiple hard photon
emission.

The radiatively corrected cross section (\ref{eq:factorization}) is
implemented in a Monte Carlo event generator by solving (\ref{eq:DGLAP})
by iteration.  This procedure is well known from the corresponding QCD
applications \cite{QCD} and, as a by-product of the algorithm, the
four-momenta of the radiated photons may be generated explicitly.  For
more details we refer the reader to \cite{Good-Stuff}.

{\tt WOPPER} also includes the effects from finite $W$ width.  To
introduce finite width for the intermediate $W$ bosons one has various
possibilities \cite{Width}.  The one chosen in the Monte Carlo is to
start from an off-shell cross section $\sigma^*$ for the process $e^+
e^- \to W^+ W^-$ with arbitrary (timelike) four-momenta $k_\pm$ of the
$W$ bosons.
\begin{equation}
  \sigma^* = \sigma^* (s; k_+^2, k_-^2)
\end{equation}
The cross section for the process $e^+ e^- \to 4$ fermions is then
obtained by convoluting $\sigma^*$ with propagators for the $W$ bosons
multiplied by the decay probability for the subsequent $W$ decay
\begin{equation}
  \sigma =
  \int \frac{ds_+}{\pi} \frac{ds_-}{\pi} \;
  \frac{\sqrt{s_+} \,\Gamma_W(s_+)}
       {(s_+ - M_W^2)^2 + s_+ \Gamma_W^2(s_+)}
  \frac{\sqrt{s_-} \,\Gamma_W(s_-)}
       {(s_- - M_W^2)^2 + s_- \Gamma_W^2(s_-)}
  \sigma^*(s; s_+, s_-)
  \label{eq:resonance-formula}
\end{equation}
where $\Gamma_W (s) \approx \sqrt{s} \, \Gamma_W/M_W$ is the effective
off-shell decay width of the $W$'s.  This means that we neglect the
contributions of the so-called background diagrams, which are suppressed
by a factor $\Gamma_W/M_W \sim$~2.5\% for each non-resonant propagator.
The contribution of the background diagrams may be reduced further by
appropriate cuts on the invariant masses of the final state.

In the Monte Carlo generator {\tt WOPPER} the four fermion final states
are generated according to the distribution
(\ref{eq:resonance-formula}).  In the partial decay widths of the $W$'s,
the QCD corrections to the hadronic decays have been taken into account
up to first order in $\alpha_S$.  The decay angular correlations of the
fermions are calculated by using the polarization density matrix of the
intermediate $W$'s, which is obtained from the off-shell helicity
amplitudes.  For more details see ref.~\cite{Good-Stuff}.


\section{Results}

Since {\tt WOPPER} is a full Monte Carlo event generator, one can in
principle study the corrections due to finite width and electromagnetic
radiation for any exclusive quantity.  However, for the sake of
comparison with other work we will consider here mainly corrections to
the total cross section and several simple distributions.

In figure~1 we plot the total cross section in the energy range from the
$W$ pair production threshold to 1~TeV.  The dotted, dashed and full
lines show the cross section obtained from {\tt WOPPER} at Born level in
the narrow width approximation, including finite $W$ width, and with all
corrections turned on, respectively.  Figure~2 gives the corrections
relative to the lowest order cross section due to finite $W$ width and
QED corrections in the region from 200~GeV to 1~TeV.  As it was pointed
out earlier \cite{BD-MAH}, the effects from the finite $W$ width do not
vanish at high energies but rather enhance the cross section by about
6\% at 1 TeV.

As has been mentioned above, the longitudinal modes of the electroweak
gauge bosons play a specific r\^{o}le in investigating the origin of
electroweak symmetry breaking and in extracting the effects of physics
beyond the SM.  Hence a determination of the $W$ helicities bosons from
the decay products is mandatory.  Analysises of this type have been
performed for all $W$ decay channels but without QED corrections
in~\cite{EE500}.

The polarizations of the $W$'s may be determined from their decay
angular distribution in their rest frame, which is obtained from the
energy spectrum of the decay fermions.  E.g.\ for the leptonic decays,
one defines the decay angle $\cos\theta^*$ by:
\begin{equation} \label{eq:costhstar}
  \cos\theta^* = \frac{1}{\beta}
  \left( \frac{2 E_{Lepton}}{E_{Beam}} - 1 \right)
  \qquad \mbox{ where }
  \beta = \sqrt{1 - \left(\frac{M_W}{E_{Beam}}\right)^2}
\end{equation}
Here $E_{Lepton}$ is the energy of the charged lepton in the lab system,
and $\beta$ is the $W$ velocity.  Figure~3 shows the corresponding
distribution reconstructed from the decay leptons for a total of $10^5$
events generated at 500~GeV.  One can easily see that the QED
corrections heavily distort this angular distribution and may therefore
make the determination of the $W$ helicities difficult.  In particular,
the longitudinal components of the $W$ bosons seem to be strongly
enhanced; this is due to the large boost between lab and c.m.\ system
when hard photons are radiated from the initial state.  A simple
reconstruction scheme, which accounts for this boost effect for the
special case of semi-leptonic events but assuming the narrow width
approximation, has been presented in~\cite{Hawaii}.

Finally, since {\tt WOPPER} is a true multiphoton Monte Carlo event
generator, we show in figure 4 the multiplicity distribution of events
with multiple registered hard photons at a c.m.s.\ energy of 500~GeV.
In this plot, a photon is counted if its energy lies above a given cut
$E_{\gamma,min}$ and if its polar angle lies in the range $5^\circ <
\theta_\gamma < 175^\circ$ with respect to the beam line.  For realistic
energy and angular resolutions of the detector, one expects a cross
section of the order of $5\cdot10^{-2}$pb for events with at least two
detected photons, which corresponds to about 500 events for a typical
year of running at a luminosity of $10^{33}\mbox{cm}^{-1}\mbox{s}^{-1}$.


The present version of {\tt WOPPER} does not contain weak corrections.
In a forthcoming version of the Monte Carlo generator, weak corrections
will be included in the framework of effective Born cross sections
\cite{DBD92:FKK+92}.  Also, an interface to hadronization Monte Carlos
\cite{AKV89} for a realistic description of the hadronic decays of the
$W$'s will be added.  Anomalous couplings for the $\gamma WW$ and $ZWW$
vertices may also be included in a future version of the generator.



\section*{Figure Captions}

\begin{description}
\item[Figure 1:] Total cross section from {\tt WOPPER} (full line: fully
  corrected, dashed line: finite width only, dotted line: Born formula).
\item[Figure 2:] Corrections relative to the Born cross section in the
  energy range from 200~GeV to 1~TeV (stars: fully corrected, open
  symbols: finite width only).
\item[Figure 3:] Decay angle of $W$'s into charged leptons, as given by
  eq.(\ref{eq:costhstar}), for $10^5$ events at 500~GeV.
\item[Figure 4:] Cross section for multiphoton events at 500~GeV.
  Photons are counted above $E_{\gamma,min}$ and in the angular range
  $5^\circ < \theta < 175^\circ$.
\end{description}


\unitlength1mm

\begin{figure}[p]
  \begin{picture}(160,200)

    \put(-10,110){\makebox(80,80){\epsfxsize=8cm \leavevmode
        \epsffile{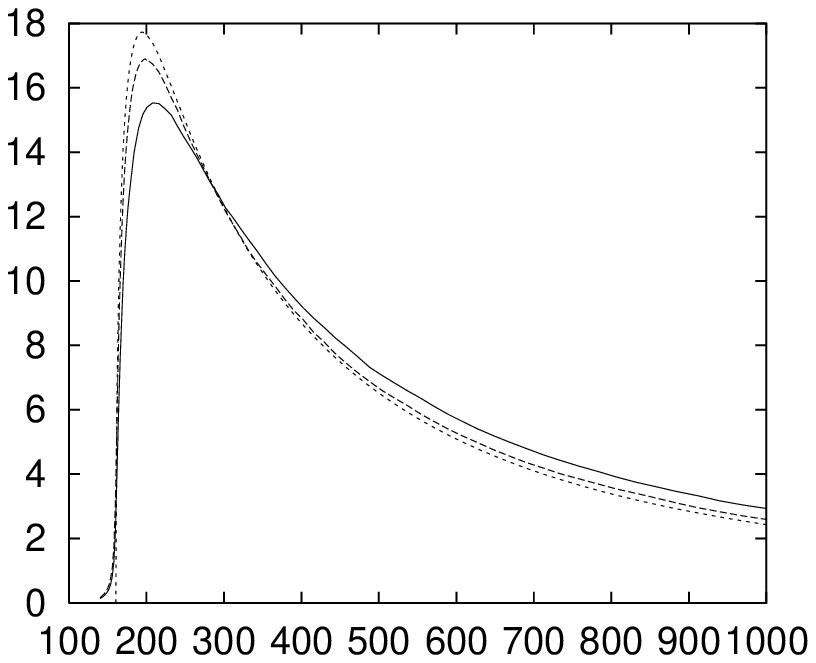}}}
    \put(-2,182){$\sigma$ [pb]}
    \put(50,116){$\sqrt{s}$ [GeV]}
    \put(30,105){Figure 1}

    \put(75,110){
      \makebox(80,80){\epsfxsize=8cm \leavevmode
        \epsffile{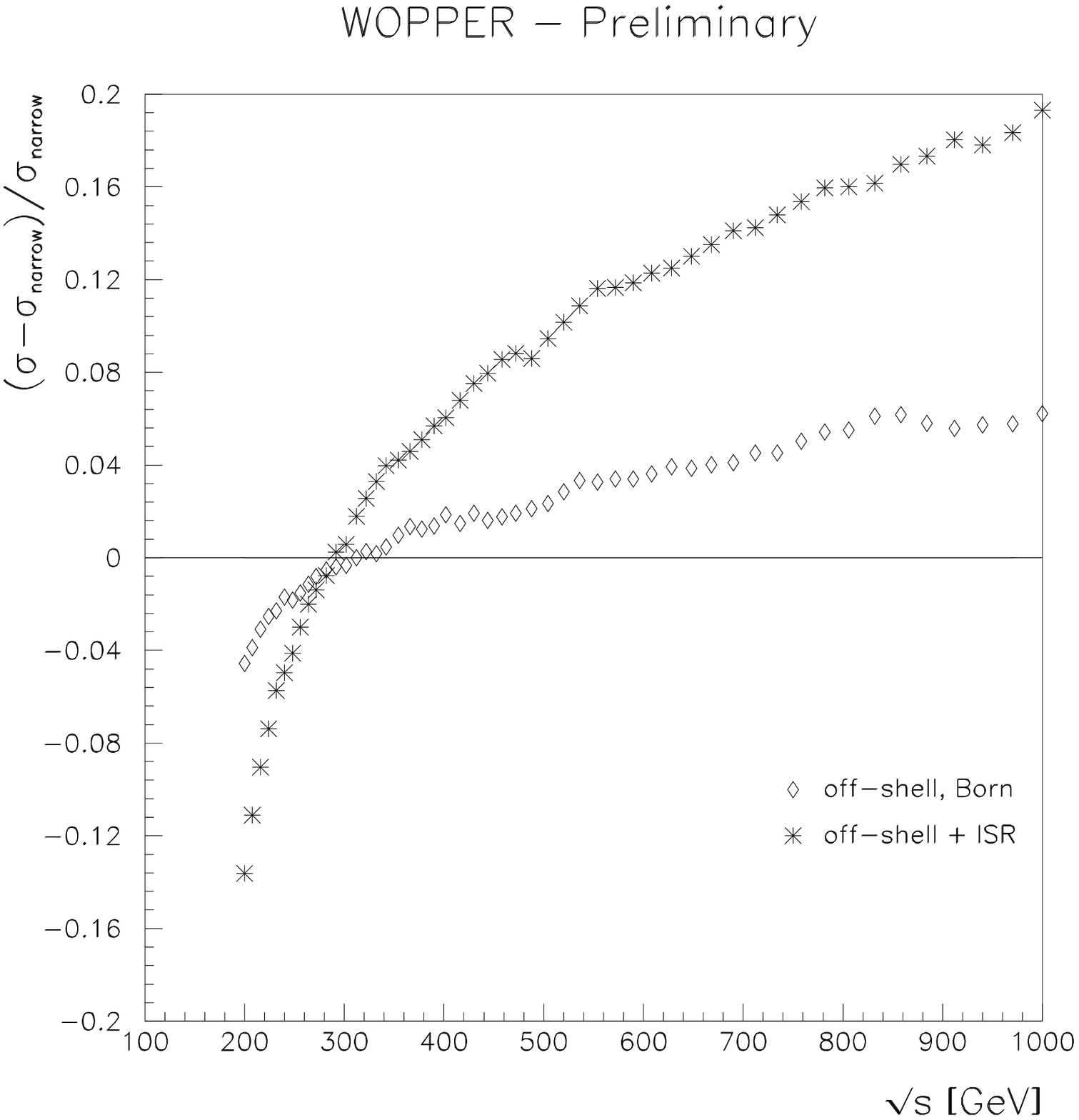}}}
    \put(110,105){Figure 2}

    \put(-5,5){\makebox(80,80){\epsfxsize=8cm \leavevmode
        \epsffile{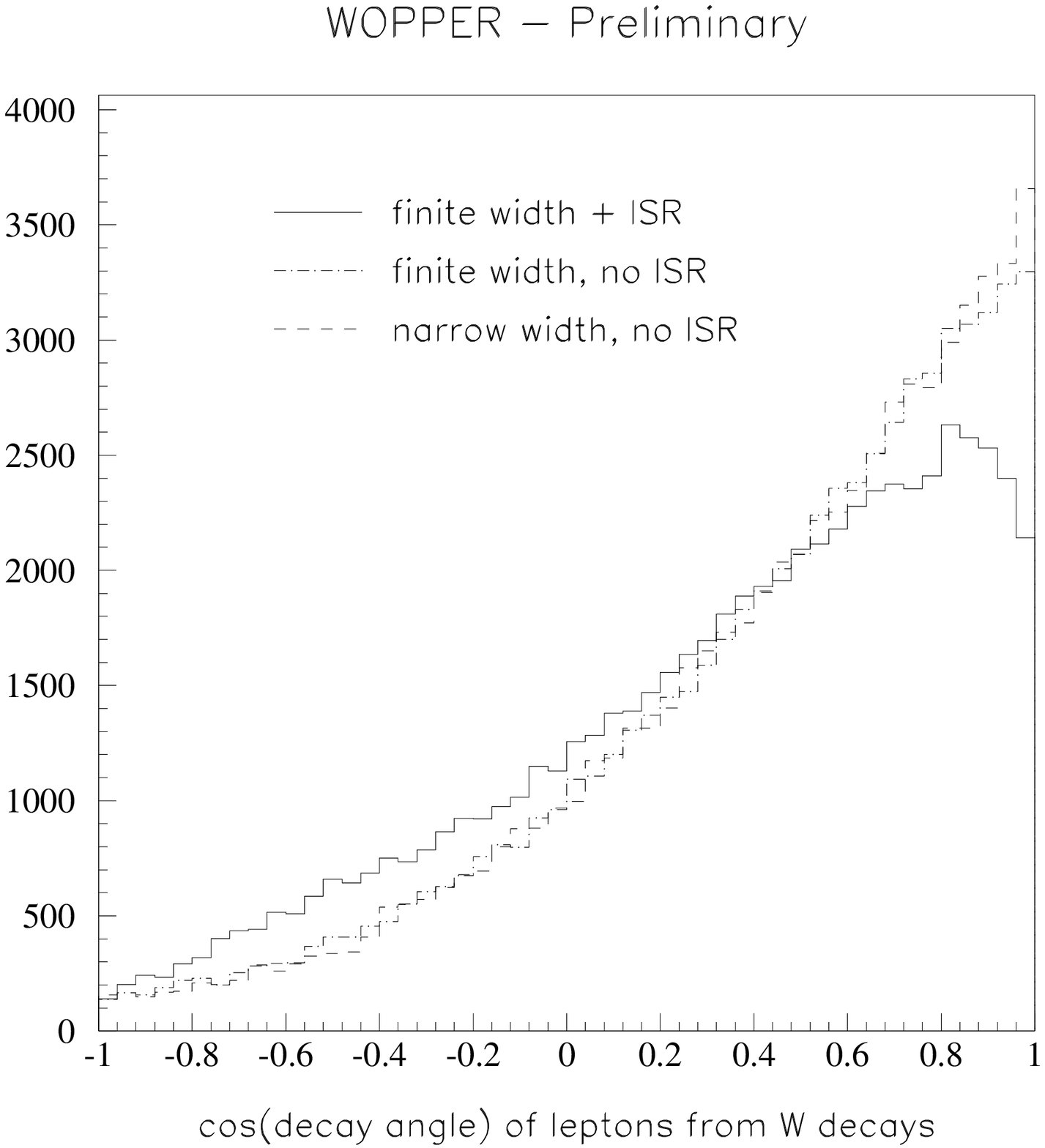}}}
    \put(30,-5){Figure 3}

    \put(75,5){
      \makebox(80,80){\epsfxsize=8cm \leavevmode
        \epsffile{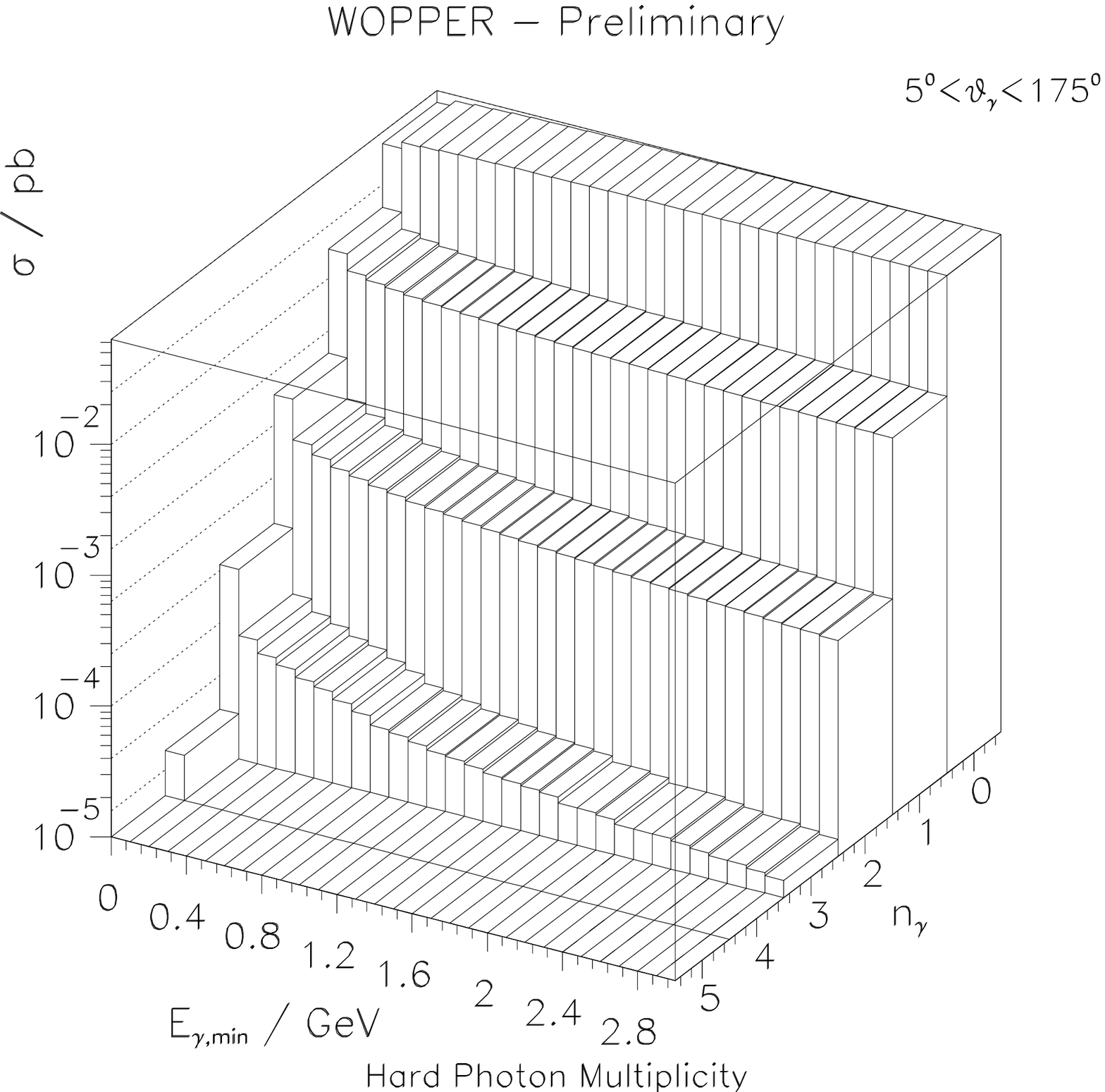}}}
    \put(110,-5){Figure 4}

  \end{picture}
\end{figure}

\end{document}